# Predictive Simulations for Tuning Electronic and Optical Properties of SubPc Derivatives

Michael J. Waters, Daniel Hashemi, Guangsha Shi, Emmanouil Kioupakis, and John Kieffer

**Abstract.**

Boron subphthalocyanine chloride is an electron donor material used in small molecule organic photovoltaics with an unusually large molecular dipole moment. Using first-principles calculations, we investigate enhancing the electronic and optical properties of boron subphthalocyanine chloride, by substituting the boron and chlorine atoms with other trivalent and halogen atoms in order to modify the molecular dipole moment. Gas phase molecular structures and properties are predicted with hybrid functionals. Using positions and orientations of the known compounds as the starting coordinates for these molecules, stable crystalline structures are derived following a procedure that involves perturbation and accurate total energy minimization. Electronic structure and photonic properties of the predicted crystals are computed using the GW method and the Bethe-Salpeter equation, respectively. Finally, a simple transport model is used to demonstrate the importance of molecular dipole moments on device performance.

**Introduction.**

Organic photovoltaics (OPV) are a possible technology for large-scale deployment of renewable energy generation. They have the advantages of being more easily processed, using less material, and being more substrate-independent than traditional inorganic PVs such as silicon.[1] Of the common OPV materials, boron subphthalocyanine chloride is unusual for having a large molecular dipole moment. As opposed to the typically planar geometry of phthalocyanines, which consist of four fused diiminoisoindole rings, boron subphthalocyanine chloride adopts an inverted umbrella shape with only three fused diiminoisoindole rings (Fig. 1). Previously, boron subphthalocyanine chloride has been shown to offer improved efficiency over phthalocyanines,[2] which previous experimental work attributes in part to molecular ordering at the donor/accepter interface creating interfacial fields. This has also been observed in polar phthalocyanines.[3]

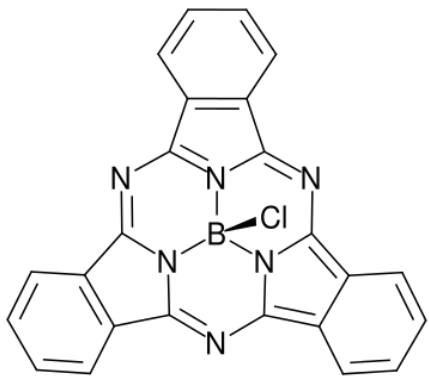

Figure 1. The structure of the boron subphthalocyanine chloride molecule, indicating the group-III metal and the halide sites.

In this work, we use *ab initio* methods to computationally evaluate and identify useful derivatives of boron phthalocyanine chloride for OPV applications, where the boron and chlorine atoms are substituted for other trivalent elements and halogen atoms, respectively. These substitutions are chosen so as to have the greatest effect on the molecular dipole moment. The trivalent site elements explored in this work are boron, aluminum, gallium, indium, scandium, and yttrium. The halogen site elements used in this work are fluorine, chlorine, bromine, and iodine. Thus, a total of twenty-four molecules are examined in our simulation matrix. For the remainder of this paper, to designate a subphtalocyanine with a given pairing of trivalent element, T and halogen element, X, we use the abbreviation T-X. For example, B-Cl refers to the commonly described boron subphthalocyanine chloride. Of the molecules explored here, the B-F, B-Cl, and B-Br molecules have been observed experimentally and their crystal structures are known.[4, 5] In order to evaluate these concept molecules as OPV materials, the structure, as well as several electrical and optical properties, are calculated for each molecule both in the gas phase and in their predicted crystalline solid phases.

Many of the molecules simulated in this work have thus far eluded synthesis,[4, 34] Therefore, by identifying the most promising candidates for a given application via the simulation-based predictive evaluation of molecular concepts, we hope to facilitate on-target materials selection and provide motivation for the development of the necessary synthesis routes.

The B-Cl molecule is used to validate our electronic and optical properties calculations since its properties are known experimentally. The B-Cl compound forms a purple material with an optical bandgap of $2.0 - 2.1$ eV.[2, 4, 6] The B-CL molecule also has interesting non-linear optical properties.[7]

The crystal structures of the B-F, B-Cl, and B-Br derivatives have been found using X-ray diffraction to be orthorhombic of space group *Pnma*, with each unit cell containing four molecules (Fig. 2).[5, 8] All molecules have the shape of an inverted umbrella with a molecular dipole pointing from the electronegative halogen towards trivalent site. In the crystalline phase, there is no net polarization due to the alternating arrangement of molecular dipoles.

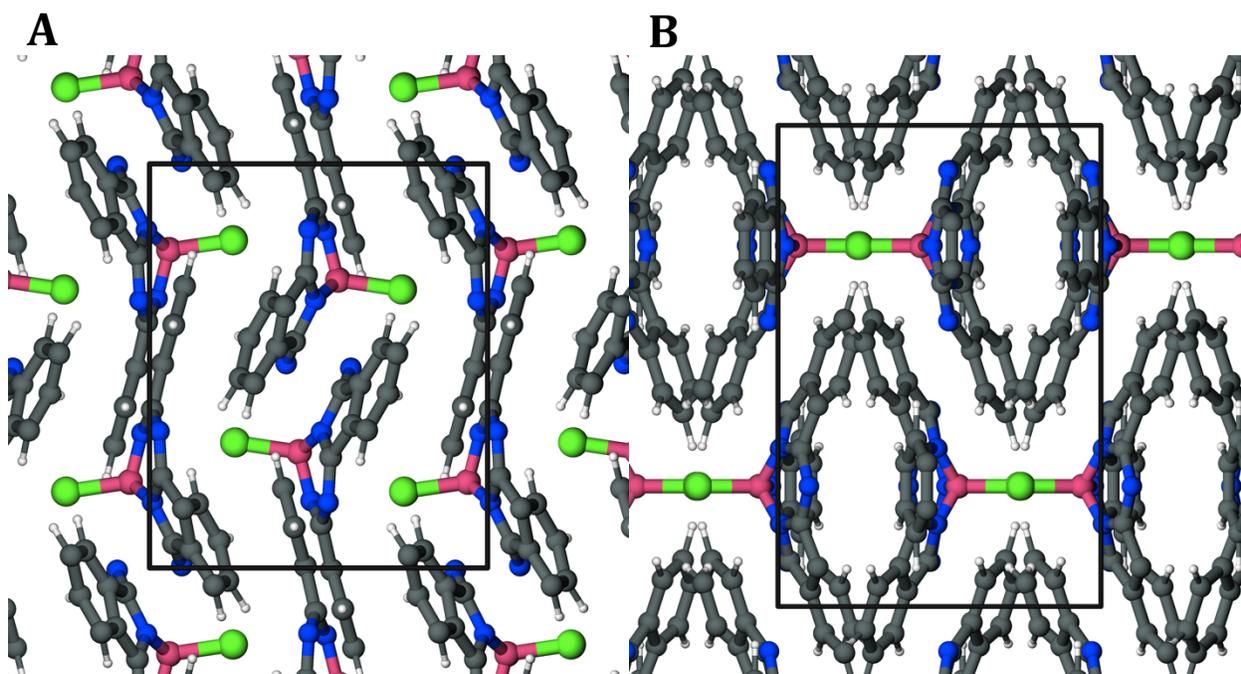

Figure 2. The B-Cl unit cell viewed along the *c* direction (left) and *b* direction (right) rendered from Ref. 8. In both renderings, the *a* direction is horizontal.

**Properties of interest.**

For the gas-phase molecules, the relaxed geometry is used to compute the HOMO/LUMO relative to the vacuum level. These levels are crucial for band alignment in organic electronic device design.[9] The molecular dipole moments were also calculated from the relaxed geometry. The dipole moment of a molecule can play a significant role in electronic properties of interfaces.[10]

To study the electronic and optical properties of SubPC derivatives in their solid phase, we need to predict the lowest-energy crystal structure, which is a difficult and outstanding problem in solid-state physics and materials science.[11, 12] Since the geometries of the isolated molecules do not considerably differ from each other, we obtain the structures of the predicted molecules via energy relaxation starting from the experimentally known structures of the synthesized derivatives. This method is described in more detail in the methodology section.

Once possible crystal structures are found, the properties of greatest interest for OPV applications are the electronic band structure and the refractive indices.[13] Additionally, the lowest-energy peak of the extinction coefficient corresponds to the optical band gap and allows the calculation of exciton binding energy. The static permittivity is also calculated using the long-wavelength limit of the complex permittivity. The static permittivity is often used in modeling exciton dissociation energy such as:[9]

$$E_{e+h} = \frac{q^2}{4\pi\epsilon_o\epsilon_r R_{e+h}},$$

where $q$ is the electron charge, $\epsilon_r$ is the relative permittivity, $\epsilon_o$ is the permittivity of free space, and $R_{e+h}$ is an effective separation distance between the electron and hole.

**Computational methodology.**

To obtain a starting point for the isolated molecular geometry, molecular structures are created and relaxed using the program Avogadro with its built-in empirical classical potentials.[14] The relaxed molecular geometry of isolated molecules is computed via relaxation using DFT and Gaussian 09 (Rev. C). The exchange-correlation functional employed is B3LYP, which is a variation of the hybrid functional created by Becke.[15] The B3LYP hybrid functional utilizes a linear combination of the correlation functionals from VWN[16] and LYP.[17] The basis set used for molecules with atoms below atomic number 36 is 6-31G(d) from Ref. 18. For molecules containing species with higher atomic number, the DGDZVP basis set is used.[19, 20] The dipole moments of the molecules are calculated from the volume integrals of the electron density.

Calculations of crystalline phases are done using VASP (version 5.3.3). These calculations employ the PAW method[21] using the PBE exchange-correlation functional[22] and the augmented-plane-wave basis set. The van der Waals interactions are accounted for using the VdW-DFT method developed in Refs. 23 and 24 with removed PBE correlation correction. For relaxations of the crystal structures, the electronic component is accomplished using the Kosugi algorithm.[25] A maximum energy difference between iterations of 0.01 meV serves as the convergence criterion. Partial occupancies are determined with a Gaussian smearing width of 0.1 eV. The cutoff energy for the plane wave basis set is 400 eV. The number of Fourier space grid points used is double that of the real space grid points.

**Crystal structure prediction.**

The state of the art of crystal structure prediction based on codes such as USPEX, CALYPSO, XtalOpt, and Polymorph/GULP allows for ~100 atoms per unit cell.[26] Subphthalocyanines are relatively large molecules with 44 atoms each and the known structures contain four molecules per unit cell. Hence, using any of the available crystal structure prediction codes is impractical. Instead, we devise a sequence of procedures involving manipulation of atomic positions and structural relaxation, based on the assumption that the new derivatives conceived here have crystals structures similar to those of the three known crystal structures. Additionally, achieving the same results via different structural progression pathways provides additional confidence in the predicted structure. Moreover, simple substitution and relaxation does not necessarily result in the lowest energy state structure. Of all the approaches we explore, the following three treatments produced the lowest energy structures:

In a first treatment, we simply substitute elements in the known B-Cl crystal structure. The structures and their corresponding unit cells are relaxed while maintaining their space group symmetry. The relaxed structures are then relaxed again without symmetry conservation. As a simple means of perturbing the structures, the unit cells and molecules inside are dilated by 10% in each lattice direction and then allowed to relax again. In the second and third treatments, elements are replaced in the isolated B-Cl molecules with known structure and the isolated derivative molecules are relaxed before they are assembled into a crystalline configuration. During the assembly procedure the trivalent atom in the derivatives is positioned on the boron location of the known B-Cl crystal structure and the relaxed gas-phase molecules are rotated about their trivalent site so that their halogen bonds and nitrogen bonds are aligned with the corresponding bonds in the known B-Cl crystal structure. From this point, in the second treatment the unit cell and molecular positions are expanded independently along each lattice direction by small increments. Upon each incremental expansion, a single point energy calculation is performed to determine the energy associated with this configuration. The procedure is repeated until the energy minima in each independent crystallographic direction are clearly identified. The corresponding unit cell vector dilations are then applied to the original structure and it is allowed to relax. The third treatment is essentially the same as the second with the exception that the molecules themselves are not dilated with the unit cell. The purpose of these procedures is to systematically perturb the crystal unit cells after elemental substitutions and to provide for different relaxation pathways and to verify that the equilibrium structures are reached in each case.

**Electronic structure calculations**

DFT calculations with the PBE functional significantly underestimate the electronic band gaps of semiconductors and insulators. This under-prediction is a well-known weakness of PBE. To remedy this, the GW approximation is used. The complex dielectric function is computed using the method in Ref. 27. The Bethe-Salpeter equation (BSE) is used to calculate the complex dielectric function of the materials in order to account for excitonic effects.[28]

For the GW calculations, the recommended GW PAW pseudo-potentials included with VASP are used with a cut off energy of 300 eV. The Gaussian energy smearing is set to a width of 0.05 eV. GW response functions are truncated to 100 eV since increasing this cut-off had no effect. 1216 bands with 50 frequency points are included in the calculation. Increasing the number of total bands considered does not affect the predicted extinction coefficient. A grid of 2x2x2 $k$-points is used in the reduced Brillouin zone since larger grids are computationally intractable.

BSE calculations are carried out with 256 frequency points, and the GW response functions are truncated at the same value as in the previous GW calculations. 100 occupied and unoccupied states are considered for the calculation. The complex shift in the Kramers-Kronig transformation for determining the dielectric function is set to 0.03 eV

**Gas phase results.**

After structural relaxation, all of the proposed subphthalocyanine derivatives retain the 3-fold symmetric inverted umbrella structure. The degree of openness of the umbrella structures canopy varies between the proposed derivatives. To quantify this openness, the angle between the nearest nitrogen atom, the trivalent atom, and the halogen is computed. This nitrogen-trivalent-halogen (NTX) angle depends almost entirely on the size of the trivalent atom as seen in Figure 3. This degree of openness may be useful for conformation or epitaxy to substrates.

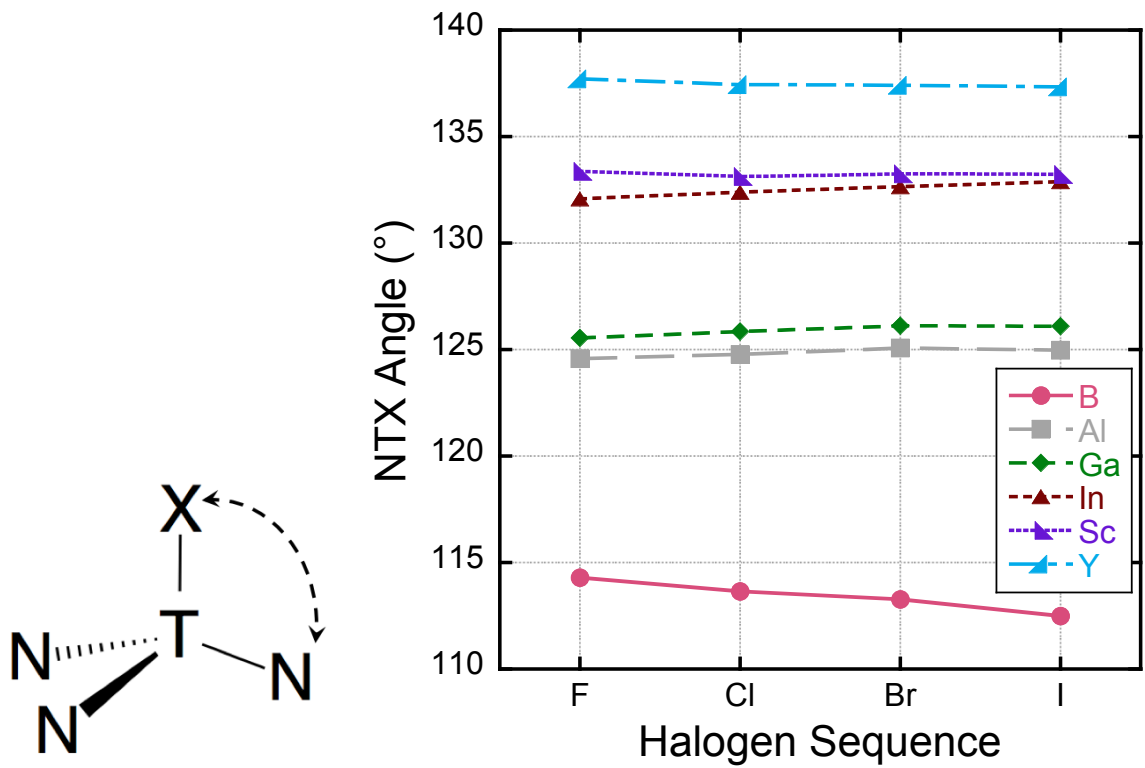

Figure 3. In the plot of the nitrogen-trivalent-halogen angle above, the nitrogen-trivalent-halogen angle depends strongly on the type of trivalent atom, but it is approximately insensitive to halogen substitution.

The molecular dipole moments of the proposed subphthalocyanine derivatives are found to vary widely from 1 ~ 6 D. All of the dipole moments are axial pointing in the direction from the halogen atom to the trivalent atom. No simple model or trends could be found to predict the variations in the molecular dipole moments as seen in Figure 4. Using electronegativity or a combination of the Mulliken charges of nearest nitrogen, trivalent, and halogen atoms and their locations do not describe the dipole moments found. Despite the lack of a model for describing the trends found, the range of dipole moments allows for a new experimental parameter to be exploited.

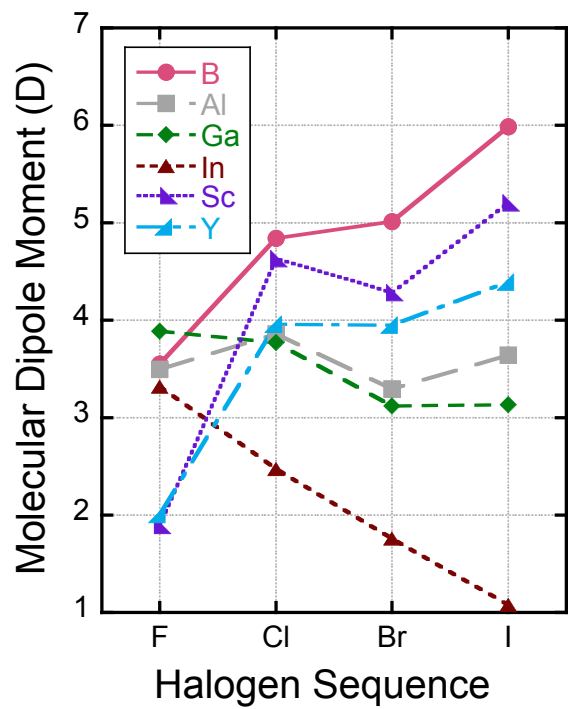

Figure 4. Chemical substitution of the trivalent metal and halogen atoms yields a broad range of molecular dipole moments .

The HOMO and LUMO levels are not strongly affected by the substitution of either halogen or trivalent species. We attribute this to the spatial distribution of the HOMO and LUMO, which are almost entirely located on the diiminoisoindole rings. This is visualized with the B-F molecule in Figure 5a. One rudimentary trend that can be observed is that with higher atomic number substitution the HOMO/LUMO gap decreases, which can be seen in, Figure 5b.

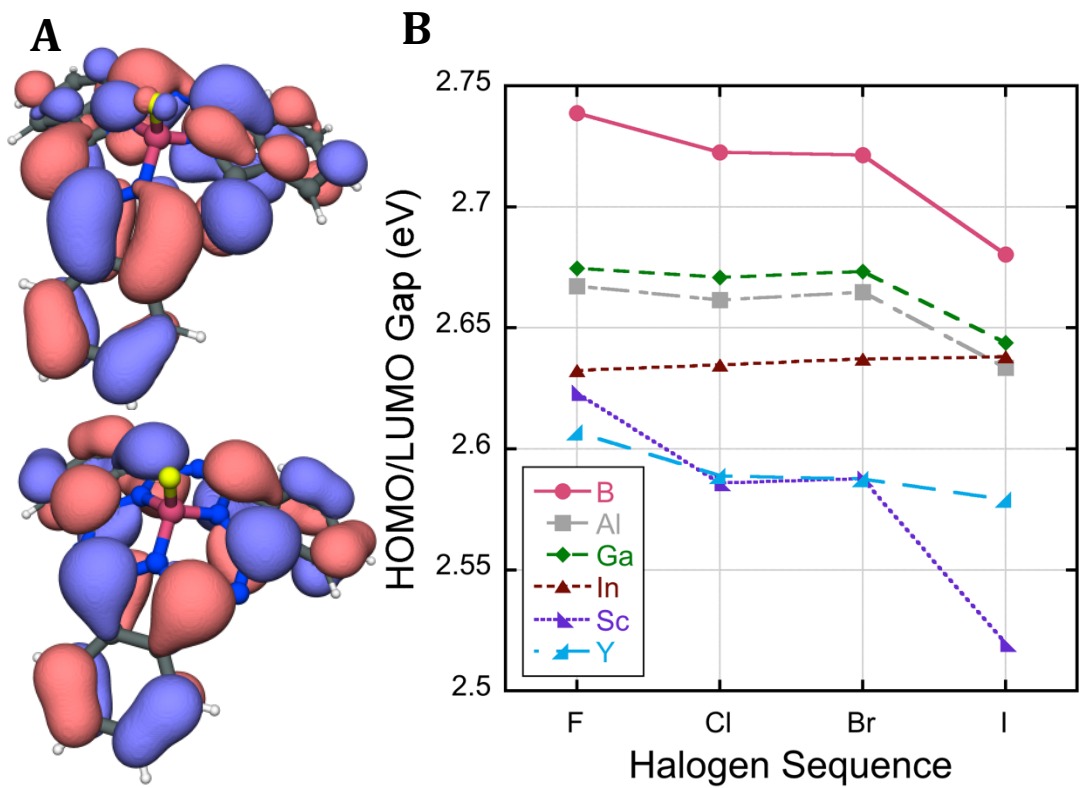

Figure 5. (A) The HOMO (bottom) and LUMO (top) wave functions of B-F. The isosurfaces contain 80% of the charge density of each state. (B) The HOMO-LUMO gap as a function of chemical substitution.

**Crystal structure results.**

The crystal structures determined for the concept molecules match experiment well for the known molecules B-F, B-Cl, and B-Br. The largest lattice parameter error is ≈ 4.5%. All of the crystal structures found are orthorhombic of the space group *Pnma*. Interestingly, the equilibrium structures of the scandium derivatives no longer have axial halogen atoms. Instead, the halogen atoms are displaced towards the second nearest scandium atom. This is apparent in Figure 6, where the Al-Cl and Sc-Cl structures are rendered side by side. Unfortunately, the crystals containing yttrium and indium derivatives are not stable when subject to systematic perturbations used to find the minimum energy structures. This may simply indicate that these molecules form different crystalline structures that cannot be accessed using the above procedure, which implicitly lends credibility to our approach in that it shows selectivity. The lattice parameters for all the proposed crystal structures are available in Table 1.

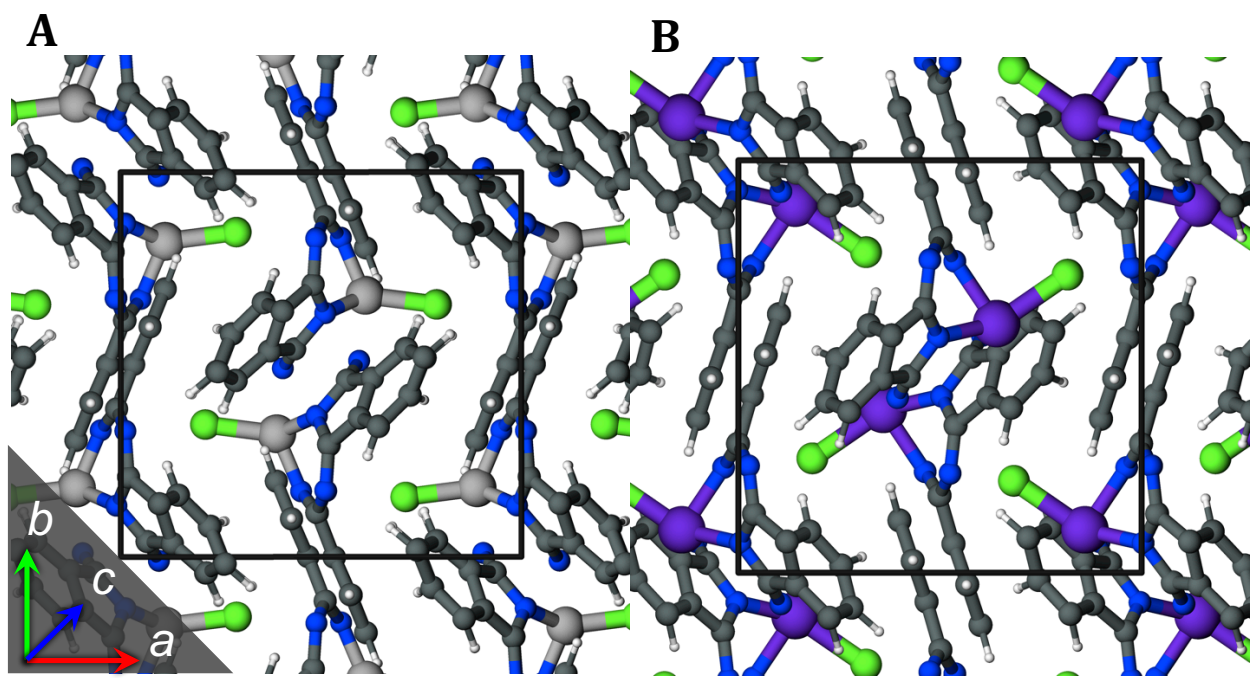

Figure 6. The crystal structure of Al-Cl (A) and Sc-Cl (B). Both unit cells are viewed such that the a, b, and c lattice directions are horizontal, vertical, and into the page respectively. The metal-halogen bonds (i.e., the electric dipole) in Al-Cl are axially aligned (A) whereas in Sc-Cl they are not (B). The metal-halogen bonds in all scandium derivatives are tilted towards the scandium atom of neighboring molecules.

Table 1. The predicted lattice parameters of the subphthalocyanine derivatives. The percentage deviations from experiments are shown in parenthesis if available.[5]

| Molecule | $a$ Lattice Parameter (Å) | $b$ Lattice Parameter (Å) | $c$ Lattice Parameter (Å) |
|---|---|---|---|
| B-F | 9.8609 (-4.5 %) | 11.9603 (-1.2 %) | 14.4059 (0.6 %) |
| B-Cl | 9.9991 (-3.7 %) | 11.9465 (-1.5 %) | 14.8851 (0.4 %) |
| B-Br | 10.1315 (-2.6 %) | 11.9319 (-0.7 %) | 15.1059 (0.3 %) |
| B-I | 10.3203 | 11.8903 | 15.5287 |
| Al-F | 11.1761 | 11.5943 | 14.1377 |
| Al-Cl | 11.6901 | 11.2454 | 14.6624 |
| Al-Br | 11.8680 | 11.2251 | 14.8772 |
| Al-I | 11.8554 | 11.2113 | 15.3509 |
| Ga-F | 11.5458 | 11.4904 | 13.9705 |
| Ga-Cl | 12.0337 | 11.1450 | 14.4848 |
| Ga-Br | 11.8222 | 11.5221 | 14.7211 |
| Ga-I | 11.8946 | 11.6867 | 15.0485 |
| Sc-F | 10.9861 | 11.5415 | 13.5916 |
| Sc-Cl | 11.2909 | 11.5687 | 14.0733 |
| Sc-Br | 11.1962 | 11.6360 | 14.3676 |
| Sc-I | 11.1179 | 11.3186 | 14.6759 |

The PBE band-structure calculations for the B-Cl crystal show that conduction band and valence band edges are relatively flat across the reduced Brillouin zone, with extrema at the Y and Γ high-symmetry k-points. The same behaviors are reflected in the GW band structure (Figure 7). The GW electronic band gaps range from 2.25 ~ 2.61 eV.

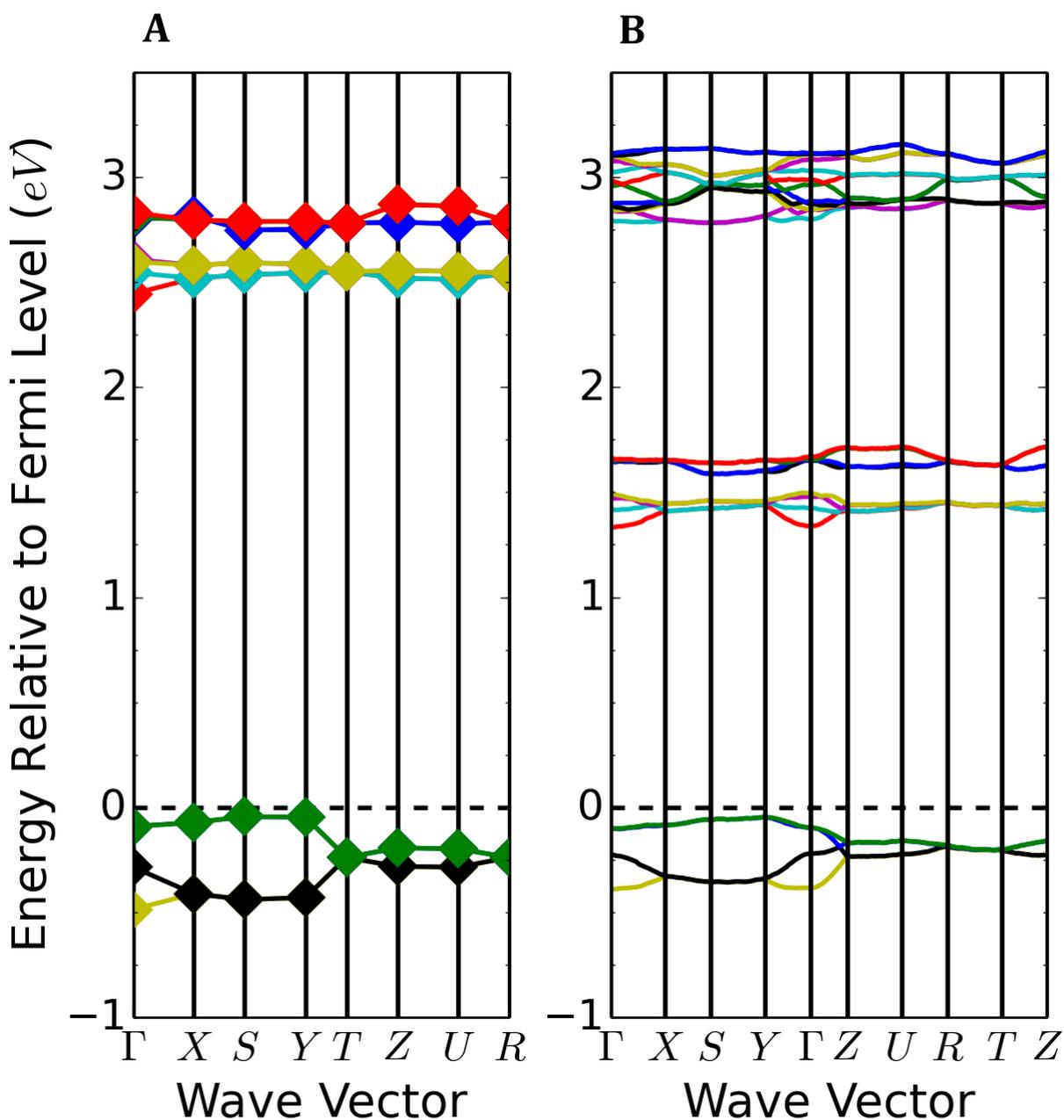

**A** **B**

Figure 7. The band structure of the B-Cl crystal calculated with (A) GW and (B) DFT. The energies are referenced to the valence band maximum. DFT underestimates the band gaps of semiconductors and insulators, but the GW method yields values in much better agreement with experiment.. Different bands are assigned different colors to guide the eye.

The complex anisotropic permittivities were computed for each crystal and are found to be highly directionally dependent. This is not unexpected considering that the similar compound, copper phthalocyanine, exhibits anisotropic permittivity relative to the molecular stacking direction.[29, 30] For the materials explored here, a relatively small complex permittivity, $\epsilon_{aa}$, is found along the $a$-direction.

Unfortunately, no single crystal optical properties measurements could be found in literature for comparison, only data for amorphous and nanocrystalline materials. Further complicating the problem is the fact that the local molecular environment can affect the absorption spectra.[31] In order to reconcile the calculations performed for anisotropic structures with measurements done on isotropic materials, the anisotropic permittivity is directionally averaged. To this end, the complex refractive index is then taken as the square root of the averaged complex anisotropic permittivity. Fig. 8 shows the directionally averaged extinction coefficient, along with those for each lattice direction and that measured for an amorphous film of the B-Cl derivative. The additional peaks in the theoretical spectra between 400~485 nm, which also occur in the other boron derivatives, do not appear in the experimental spectra for amorphous and nanocrystalline samples, and may only be observable in single crystals. These may be physical but only observable in single crystals or they are an artifact of the limited number of k-points used. Either way, the lowest-energy absorption peak position aligns well with experiment with a difference of 75 meV for B-Cl and 91 meV for B-F.[32] The lowest-energy peak positions also agree well with the extinction-coefficient measurements of Fulford *et al.* for B-Cl dissolved in toluene. The maximum difference between B-Cl, B-F, and B-Br is 31 meV.[5] These results are tabulated in Table 2.

Table 2. The electronic and optical gaps of subphthalocyanine derivatives. Experimental values are listed in parentheses where available.

| Molecule | Optical Band Gap (nm) | Electronic Band Gap (eV) | Exciton Binding Energy (eV) |
|---|---|---|---|
| B-F | 559 (583)[32] (562)[5] | 2.50 | 0.28 |
| B-Cl | 565 (585)[32] (565)[5] (587)[33] | 2.49 | 0.29 |
| B-Br | 558 (566)[5] | 2.51 | 0.29 |
| B-I | 554 | 2.58 | 0.34 |
| Al-F | 567 | 2.52 | 0.34 |
| Al-Cl | 550 | 2.54 | 0.29 |
| Al-Br | 543 | 2.56 | 0.28 |
| Al-I | 549 | 2.56 | 0.30 |
| Ga-F | 571 | 2.47 | 0.30 |
| Ga-Cl | 580 | 2.52 | 0.38 |
| Ga-Br | 546 | 2.56 | 0.29 |
| Ga-I | 540 | 2.61 | 0.31 |
| Sc-F | 621 | 2.24 | 0.25 |
| Sc-Cl | 566 | 2.37 | 0.18 |
| Sc-Br | 546 | 2.47 | 0.20 |
| Sc-I | 552 | 2.50 | 0.25 |

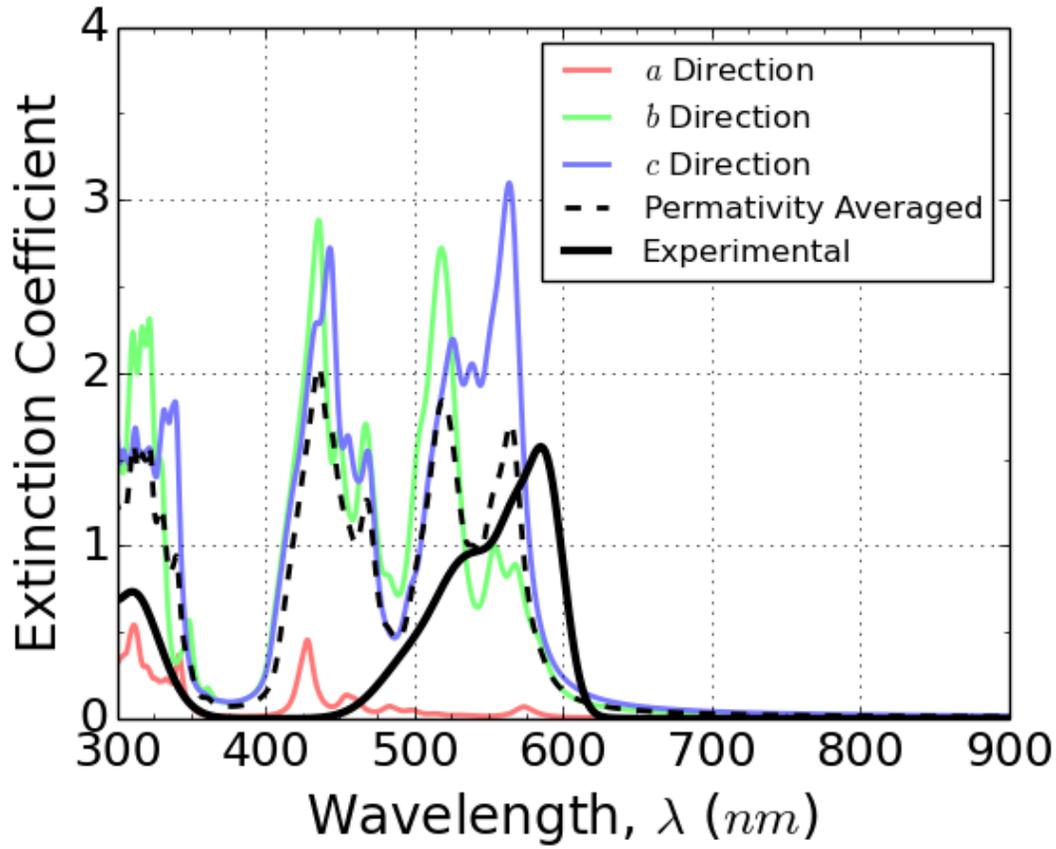

Figure 8. The calculated directionally dependent and directionally averaged extinction coefficient of B-Cl as a function of photon wavelength. The calculated optical gap matches well the available experimental data from Ref. 32. Similar predictive accuracy is expected for other investigated SubPC derivatives.

The static permittivity at the long wavelength limit is also computed with the BSE calculations. The static permittivity directly controls the exciton binding energy, which is important for OPV device performance. In the proposed derivatives, the static permittivity is found to generally decrease with heavier halogen atoms, which can be seen in Figure 9. To our knowledge, no static permittivity measurements of any of the proposed derivatives are available in the literature for validation. However, the values presented here can still be used for relative comparisons between the substituted SubPCs.

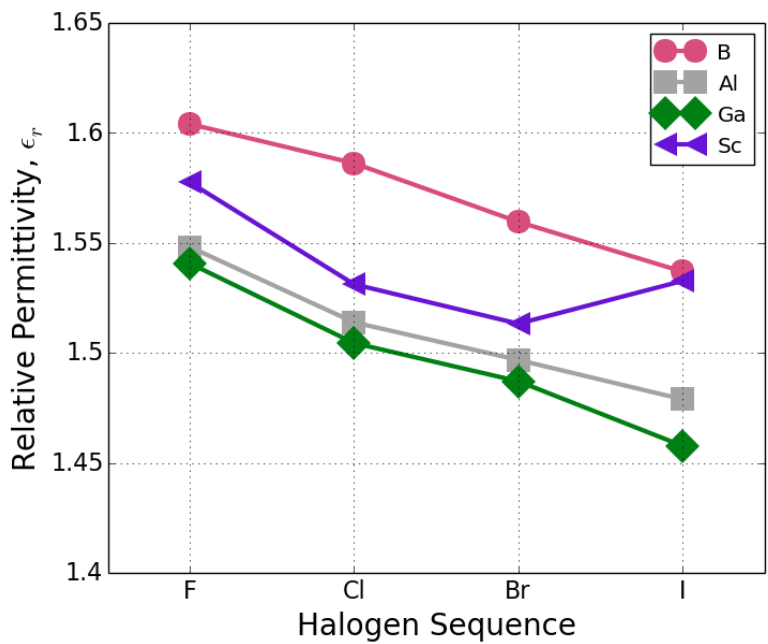

Figure 9. The static permittivity of the proposed SubPC derivatives as a function of chemical substitution.

Finally, we simulate the I-V curves for the solar cell of Ref. 32 using the methods contained therein. These simulations were carried out under the assumption that one quarter of the $C_{60}$ sites are occupied with ball-in-cup aligned subphthalocyanine derivatives. The I-V curves for four of the investigated derivatives are show in Fig. 10. These include B-I as it has the largest molecular dipole moment, In-I as it has the smallest molecular dipole, Sc-F as it has the smallest molecular dipole of the derivatives that were found to be stable enough for crystal structure prediction, and B-Cl as it is the commonly known derivative. In this simulation, B-I can offer a small improvement over B-Cl.

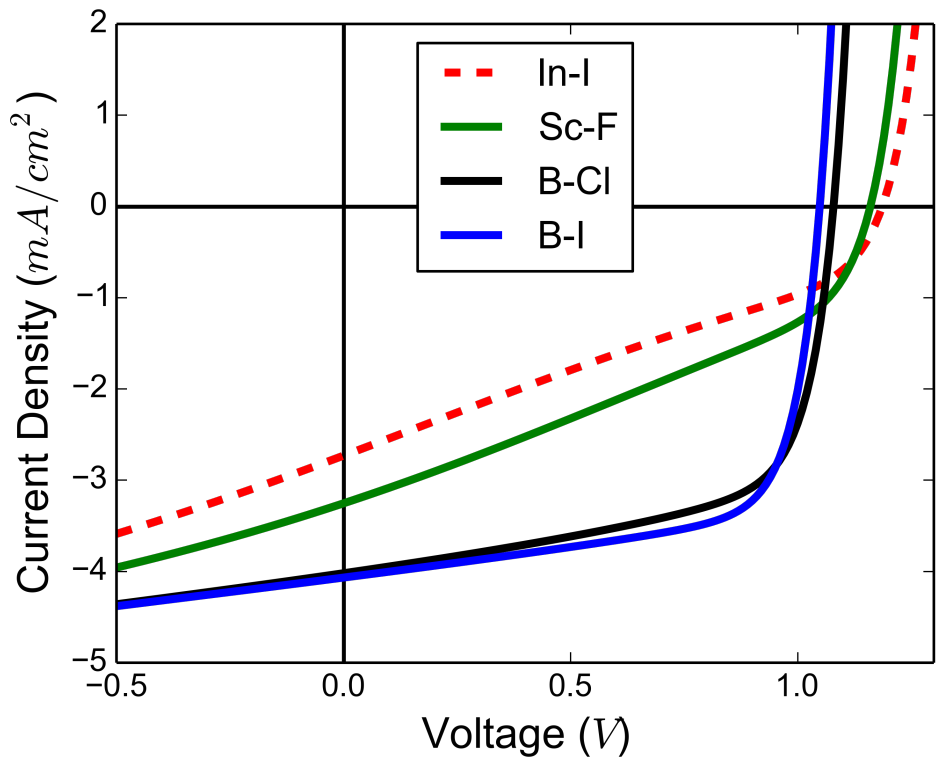

Figure 10. Simulated I-V curves for a solar-cell device substituting four of the investigated SubPC derivatives. The simulation assumes some molecular dipole ordering at the interface.

**Conclusions.**

We predicted the crystal structures, electronic band structure, and optical properties for series of SubPc molecules generated through systematic elemental substitutions. Our results show that the degree of openness of the molecule canopy is controlled by the trivalent atom. The predicted molecular dipole moments vary over a broad range from 1 ~ 6 D. All crystalline solid phases are found to be orthorhombic in the *Pnma* space group. The predicted optical band gaps and onset of optical absorption spectra are in good agreement with available experimental data. The static permittivities of the molecules are found to increase within the halogen sequence. Our calculated results serve as design parameters to increase SubPC OPV device performance.

**Acknowledgements.**

Guangsha Shi and Emmanouil Kioupakis were supported by the National Science Foundation CAREER award through Grant No. DMR-1254314.

**References.**

1. Lewis, N. S. "Toward cost-effective solar energy use." *science* 315 (5813): 798–801. (2007)

2. Mutolo, K. L., E. I. Mayo, B. P. Rand, S. R. Forrest, and M. E. Thompson "Enhanced open-circuit voltage in subphthalocyanine/C60 organic photovoltaic cells." *Journal of the American Chemical Society* 128 (25): 8108–8109. (2006)

3. Jailaubekov, A. E., A. P. Willard, J. R. Tritsch, W.-L. Chan, N. Sai, R. Gearba, L. G. Kaake, K. J. Williams, K. Leung, P. J. Rossky, and X.-Y. Zhu "Hot charge-transfer excitons set the time limit for charge separation at donor/acceptor interfaces in organic photovoltaics." *Nat Mater* 12 (1): 66–73. (2013)

4. Claessens, C. G., D. González-Rodríguez, and T. Torres "Subphthalocyanines: singular nonplanar aromatic compounds synthesis, reactivity, and physical properties." *Chemical reviews* 102 (3): 835–854. (2002)

5. Fulford, M. V., D. Jaidka, A. S. Paton, G. E. Morse, E. R. L. Brisson, A. J. Lough, and T. P. Bender "Crystal Structures, Reaction Rates, and Selected Physical Properties of Halo-Boronsubphthalocyanines (Halo= Fluoride, Chloride, and Bromide)." *Journal of Chemical & Engineering Data* 57 (10): 2756–2765. (2012)

6. Morse, G. E., and T. P. Bender "Boron Subphthalocyanines as Organic Electronic Materials." *ACS Applied Materials & Interfaces* 4 (10): 5055–5068. (2012)

7. Diaz-Garcia, M. A., and F. Agullo-Lopez… "High second-order optical nonlinearities in subphthalocyanines." *... and Laser Science ...* (1996)

8. Kietaibl, H. "Die Kristall- und Molek√⁰lstruktur eines neuratigen phthalocyanin√§hnlichen Borkomplexes." *Monatshefte f√°r Chemie* 105 (2): 405–418. (1974)

9. Rand, B. P., D. P. Burk, and S. R. Forrest "Offset energies at organic semiconductor heterojunctions and their influence on the open-circuit voltage of thin-film solar cells." *Physical Review B* 75 (11): 115327. (2007)

10. Campbell, I. H., S. Rubin, T. A. Zawodzinski, J. D. Kress, R. L. Martin, D. L. Smith, N. N. Barashkov, and J. P. Ferraris "Controlling Schottky energy barriers in organic electronic devices using self-assembled monolayers." *Physical Review B* 54 (20): R14321. (1996)

11. Wang, Y., J. Lv, L. Zhu, and Y. Ma "CALYPSO: A method for crystal structure prediction." *Computer Physics Communications* 183 (10): 2063–2070. (2012)

12. Oganov, A. R., and C. W. Glass "Crystal structure prediction using ab initio evolutionary techniques: Principles and applications." *The Journal of chemical physics* 124 (24): 244704. (2006)

13. Pettersson, L. A. A., L. S. Roman, and O. Inganäs "Modeling photocurrent action spectra of photovoltaic devices based on organic thin films." *Journal of Applied Physics* 86 (1): 487–496. (1999)

14. Hanwell, M. D., D. E. Curtis, D. C. Lonie, T. Vandermeersch, E. Zurek, and G. R. Hutchison "Avogadro: an advanced semantic chemical editor, visualization, and analysis platform." *Journal of cheminformatics* 4 (1): 1–17. (2012)

15. Becke, A. D. "Density-functional thermochemistry. III. The role of exact exchange." *The Journal of Chemical Physics* 98 5648. (1993)

16. Vosko, S. H., L. Wilk, and M. Nusair "Accurate spin-dependent electron liquid correlation energies for local spin density calculations: a critical analysis." *Canadian Journal of Physics* 58 (8): 1200–1211. (1980)

17. Lee, C., W. Yang, and R. G. Parr "Development of the Colle-Salvetti correlation-energy formula into a functional of the electron density." *Physical Review B* 37 (2): 785. (1988)

18. Petersson, G. A., and M. A. Al-Laham "A complete basis set model chemistry. II. Open-shell systems and the total energies of the first-row atoms." *The Journal of chemical physics* 94 6081. (1991)

19. Sosa, C., J. Andzelm, B. C. Elkin, E. Wimmer, K. D. Dobbs, and D. A. Dixon "A local density functional study of the structure and vibrational frequencies of molecular transition-metal compounds." *The Journal of Physical Chemistry* 96 (16): 6630–6636. (1992)

20. Godbout, N., D. R. Salahub, J. Andzelm, and E. Wimmer "Optimization of Gaussian-type basis sets for local spin density functional calculations. Part I. Boron through neon, optimization technique and validation." *Canadian Journal of Chemistry* 70 (2): 560–571. (1992)

21. Blöchl, P. E. "Projector augmented-wave method." *Physical Review B* 50 (24): 17953. (1994)

22. Perdew, J. P., K. Burke, and M. Ernzerhof "Generalized gradient approximation made simple." *Physical review letters* 77 (18): 3865. (1996)

23. Klimeš, J., D. R. Bowler, and A. Michaelides "Chemical accuracy for the van der Waals density functional." *Journal of Physics: Condensed Matter* 22 (2): 022201. (2010)

24. Klimeš, J., D. R. Bowler, and A. Michaelides "Van der Waals density functionals applied to solids." *Physical Review B* 83 (19): 195131. (2011)

25. Kosugi, N. "Strategies to vectorize conventional SCF-CI algorithms." *Theoretica chimica acta* 72 (2): 149–173. (1987)

26. Wang, Y., J. Lv, L. Zhu, and Y. Ma "Crystal structure prediction via particle-swarm optimization." *Physical Review B* 82 (9): 094116. (2010)

27. Gajdoš, M., K. Hummer, G. Kresse, J. Furthmüller, and F. Bechstedt "Linear optical properties in the projector-augmented wave methodology." *Physical Review B* 73 (4): 045112. (2006)

28. Onida, G., L. Reining, and A. Rubio "Electronic excitations: density-functional versus many-body Green's-function approaches." *Reviews of Modern Physics* 74 (2): 601. (2002)

29. Gordan, O. D., M. Friedrich, and D. R. T. Zahn "The anisotropic dielectric function for copper phthalocyanine thin films." *Organic electronics* 5 (6): 291–297. (2004)

30. Debe, M. K. "Variable angle spectroscopic ellipsometry studies of oriented phthalocyanine films. II. Copper phthalocyanine." *Journal of Vacuum Science & Technology A* 10 (4): 2816–2821. (1992)

31. Heremans, P., D. Cheyns, and B. P. Rand "Strategies for increasing the efficiency of heterojunction organic solar cells: material selection and device architecture." *Accounts of chemical research* 42 (11): 1740–1747. (2009)

32. S.E., Morris, Bilby. D., S. M.E., H. Hashemi., Waters M.J., Kieffer J., K. J., and Shtein. M. "Effect of axial halogen substitution on the performance of subphthalocyanine based organic photovoltaic cells." *Organic Electronics* 15 (12): 3660 – 3665. (2014)

33. Gommans, H. H. P., D. Cheyns, T. Aernouts, C. Girotto, J. Poortmans, and P. Heremans "Electro-Optical Study of Subphthalocyanine in a Bilayer Organic Solar Cell." *Advanced functional materials* 17 (15): 2653–2658. (2007)

34. Torres, T. "From Subphthalocyanines to Subporphyrins," *Angewandte Chemie International Edition* 45 (18): 2834-2837. (2006)